\documentclass[%
reprint,
amsmath,amssymb,
aps,
longbibliography
]{revtex4-2}

\usepackage{graphicx}
\usepackage{dcolumn}
\usepackage{bm}
\usepackage{hyperref}
\usepackage[dvipsnames]{xcolor}

\newcommand{\pd}[2]{\frac{\partial #1}{\partial #2}}
\newcommand{\rv}[1]{\textcolor{black}{#1}}

\begin{document}

\preprint{APS/123-QED}

\title{Nucleation suppression by charge screening on grain boundaries: a kinetic model for bulk imprint in polycrystalline ferroelectric thin films}

\author{Huanhuan Tian}
\affiliation{%
 Zhangjiang Laboratory, Shanghai, China
}%
\author{Jianguo Yang}
\email{yangjg@zjlab.ac.cn}
\affiliation{%
 Zhangjiang Laboratory, Shanghai, China
}%
\author{Ming Liu}
\email{liuming@fudan.edu.cn}
\affiliation{%
Frontier Institute of Chip and System, Fudan University, Shanghai, China
}%
\affiliation{%
 Institute of Microelectronics, Chinese Academy of Sciences, Beijing, China. 
}

\date{\today}

\begin{abstract}
The imprint effect, a significant reliability challenge in ferroelectric memories, manifests as a shift in the coercive field during retention and endurance tests, ultimately degrading the usable memory window. \rv{While traditional models attribute imprint primarily to charge screening at the interface between the dead layer and the ferroelectric film, the contribution from grain boundaries has been largely overlooked. This work advances a bulk imprint mechanism by establishing a phase-field model, which demonstrates that the tuning of domain nuclei near grain boundaries via charge screening consistently explains the imprint process and aligns with key experimental trends.}  These findings provide novel insights into the imprint process and advance the understanding of reliability issues in ferroelectric memory devices.

\end{abstract}

\maketitle

\section{Introduction}
The imprint effect, characterized by the difficulty in switching a polarization state after an extended idle time, represents a major reliability challenge in ferroelectric memories\cite{tagantsev_nature_2004, takada_timedependent_2021, silva_roadmap_2023}. It can lead to shifted coercive field and reduced memory window in both retention tests \cite{toprasertpong_low_2022} and endurance tests \cite{schenk_complex_2015}.   Recent studies suggest that the alternating application of retention and endurance stresses poses the most severe scenario for degrading the usable memory window over the lifetime of ferroelectric memory devices \cite{ettisserry_comprehensive_2024, su_root_2025}. Therefore, a thorough understanding of the imprint mechanism is essential for improving device reliability. 


\rv{While numerous experimental studies exist, this theoretical work focuses on the observations of imprint phenomena in two representative recent works \cite{kim_imprint_2024, vishnumurthy_ferroelectric_2024}: (1) Logarithmic law: the shift of the coercive field $\Delta E_c$ usually follows $\Delta E_c = E_0 \ln(1 + t/t_0)$. (2) Thermal acceleration: $E_0$ increases and $t_0$ decreases as temperature increases. (3) Asymmetric branch shifting: the shift of the two branches of the hysteresis loop is usually asymmetric. $|\Delta E_c^+|>|\Delta E_c^-|$ for imprint of the negative polarization state, and vice versa. }


\rv{The imprint phenomenon is commonly explained by an interface charge screening (ICS) model \cite{grossmann_interface_2002, grossmann_interface_2002-1, tagantsev_nature_2004}, as schematically illustrated in Fig.1(a). The depolarization field induces charge accumulation (with charge density $\sigma$) at the interface between the dead layer and the ferroelectric film, thereby pinning the polarization state. An ICS model is usually composed of Gauss's Law, Kirchhoff's Voltage Law, a dynamic charge migration model (through the dead layer \cite{tagantsev_nature_2004} or the ferroelectric film \cite{su_root_2025}), and a ferroelectric switching model (such as Nucleation-Limited Switching (NLS) model \cite{kim_imprint_2024, xu_physics-based_2025} or polycrystal phase-field model \cite{alhadalahbabi_investigating_2025}). In a compact model which assumes uniform polarization $P_r$, the electric fields in the ferroelectric film ($E_f$)  can be derived \cite{xu_physics-based_2025}:
\begin{equation}
    E_f = \frac{C_i V  - P_r + \sigma}{(C_i + C_f) h_f}, \  C_i = \frac{1}{\frac{h_i}{\epsilon_0 \epsilon_i} + \frac{\delta_{sc}}{\epsilon_0}}, \  C_f = \frac{\epsilon_0 \epsilon_{f}}{h_f} 
\end{equation}
where $V$ is the applied voltage,  $\epsilon_0$ is the vacuum permittivity, $\epsilon_i$ and $\epsilon_f$ are the dielectric constant, $h_i$ and $h_f$ are the film thickness, and the subscript $i$ and $f$ represent the dead layer and the ferroelectric film, respectively, and $\delta_{sc}$ is the total finite charge screening length of metal electrodes. The third term in $E_f$ represents $\Delta E_c$ due to ICS. }

\begin{figure}
    \centering
    \includegraphics[width=0.95\linewidth]{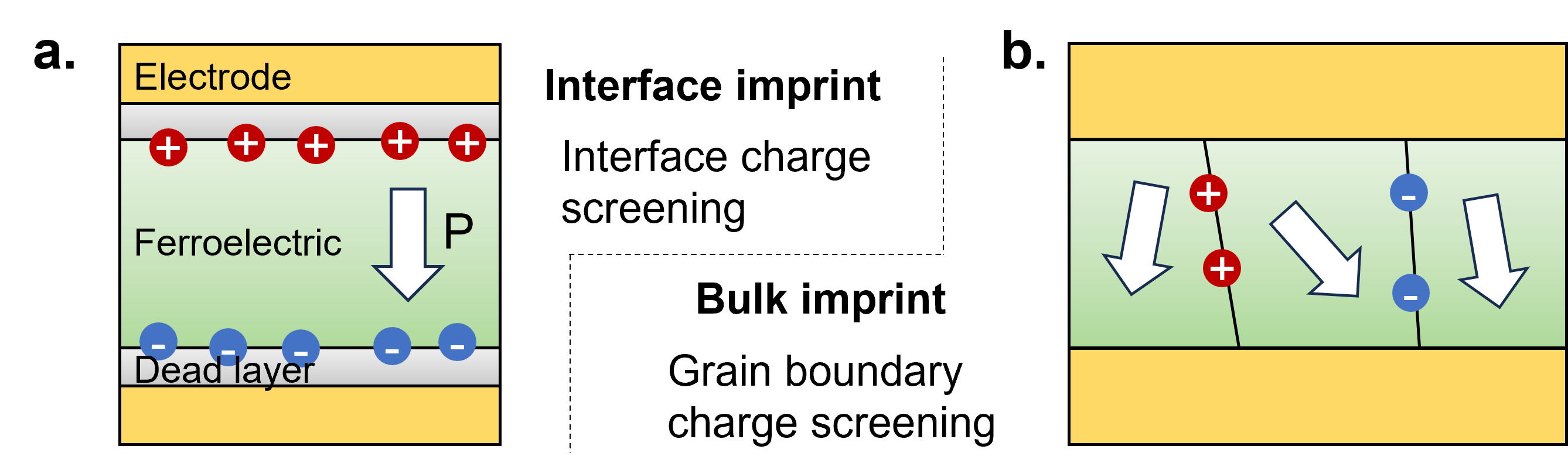}
    \caption{\rv{Schematics for imprint induced by (a) interface charge screening, and (b) grain-boundary charge screening.}  }
    \label{fig1}
\end{figure}

\rv{While most ICS models successfully capture the logarithmic law and thermal acceleration effects, they generally predict symmetric branch shifting. To address asymmetric imprint, two recent models have been proposed. Kim et al.\cite{kim_imprint_2024} attribute the asymmetry to the independent evolution of the activation field $E_a$ for domain nucleation in each branch, though the physical origin of this difference remains unclear.  Alhada-Lahdabi et al. \cite{alhadalahbabi_investigating_2025} explain it via rapid imprint recovery dynamics—driven by the fast recombination of electrons and oxygen vacancies—occurring in the first half of the hysteresis loops. }

\rv{This work proposes that grain boundary charge screening (GBCS) can also play an important role in imprint. In polycrystal ferroelectric films, uncompensated bound charges at grain boundaries generate intense local electric fields ($\sim$MV/cm), which unavoidably assist domain nucleation and attract opposite trap charges. We thus propose that the coercive field is governed by domain nucleation at grain boundaries, attributing the coercive field shift during retention to the tuning of domain nuclei by GBCS. While other ``bulk imprint" mechanisms emphasizing dipole-defect interaction within the film have been proposed before \cite{warren_voltage_1995, lee_analysis_2023}, our work exclusively identify grain boundaries as the relevant bulk defects and establish a phase-field model for the imprint process. 
The model is consistent with the three experimental trends.  }


\section{Bulk imprint model}


We write the total free energy of the ferroelectrics as \cite{tian_intrinsic_2025, wang_understanding_2019,  tian_depolarization_2025}:
\begin{equation}
    F = \int{ (f_{\rm{Landau}} + f_{\rm{grad}} + f_{\rm{elec}}) }dV,
\end{equation}
where $f_{\rm{Landau}}$ is the Landau energy, $f_{\rm{grad}}$ is the gradient energy, and $f_{\rm{elec}} = -\mathbf{P}\cdot \mathbf{E} - \frac{1}{2}\epsilon_0 \epsilon_{r,b} |\mathbf{E}|^2 
 + \rho_e \phi$ is the electrostatic energy. 
Here $\mathbf{P}$ is the polarization vector, $\mathbf{E} = -\nabla \phi$ is the electric field, $\epsilon_0$ is the vacuum permittivity, $\epsilon_{r,b}$ is the background relative permittivity, and $\rho_e$ is the free charge density. \rv{As a concept model, we ignore the elastic energy and assume $ f_{\mathrm{Landau}} = \frac{1}{2} a (P_{x'}^2 -2 P_{y'}^2 ) + \frac{1}{4}b P_{x'}^4$, where $x'$ and $y'$ are local coordinates in each grain. This $f_{\mathrm{Landau}}$ gives two stable polarization states with $P_r = \pm \sqrt{-\frac{a}{b}}$ and isotropic ionic permittivity that equals $ -\frac{1}{2a}$ \cite{tian_intrinsic_2025, tian_depolarization_2025}.}
Then the dynamics of ferroelectric switching can be described by  \cite{wang_understanding_2019}
\begin{equation}
    \frac{\tau}{\epsilon_0}\pd{\mathbf{P}}{t} = -\frac{\delta F}{\delta \mathbf{P}}  \text{ and }  \nabla\cdot (\mathbf{P} + \epsilon_0 \epsilon_{r,b} \mathbf{E}) = \rho_e,
    \label{eq:pde}
\end{equation}
where $\tau$ is a time constant. Using the depletion approximation for charge carriers, we have 
\begin{equation}
    \rho_e = e(N_D^+ - N_A^-),
\end{equation}
where $e$ is the electron charge, $N_D^+$ and $N_A^-$ are the concentration of ionized donors (D) and acceptors (A). 
The kinetics of the ionization process can be described by
\begin{equation}
    \tau_t \pd{N_D^+}{t} = (N_t - N_D^+) \exp(\frac{-e\phi - \Delta E_t}{k_b T}) - N_D^+,
\end{equation}
\begin{equation}
    \tau_t \pd{N_A^-}{t} = (N_t - N_A^-) \exp(\frac{e\phi - \Delta E_t}{k_b T}) - N_A^-,
\end{equation}
where $\tau_t$ is the characteristic trapping/detrapping time constant, $N_t$ is the number density of donor/acceptor traps, $k_b$ is the Boltzmann constant, $T$ is the temperature, \rv{$\Delta E_t$ is the effective trap energy level ($\Delta E_t  = E_F - E_D + k_bT\ln g_D = E_A - E_F + k_bT \ln g_A$, where $E_A$ and $E_D$ are trap energy levels at $\phi = 0$, $E_F$ is the Fermi level, and $g_A$ and $g_D$ are degeneracy factors). These two equations are adapted from classical formulas \cite{sze_physics_2007} by imposing the charge neutrality condition at $\phi=0$.}



To parameterize the ferroelectric model, we take $a = -2.27\times 10^9\ \rm{V\cdot m/C}$, $b = 9.09\times 10^9 \ \rm{V\cdot m^5/C^3}$, $\epsilon_{r,b} = 5$ based on HZO properties \cite{tian_depolarization_2025}. We assume $\tau_t = 10^9 \tau$ and ignore charge trapping when simulating hysteresis loops. We use the vorogoni principle to generate $N_g$ grains (mean grain size $h_g = L/N_g$) and assign them with random orientations in a region with length $L = 500\ \rm{nm}$ and height $h_f$ (which varies). For each parameter set, we simulate 5 different random polycrystalline structures.
We discretize the equations using the finite volume method with mesh size equal to the lattice size (0.5 nm) \cite{tian_intrinsic_2025}. 

\section{simulation results}

We first present results for a base case, with $h_f = 8\ \rm{nm}$, $L = 500\ \rm{nm}$,  $N_g = 25$, \rv{$\Delta E_t/k_b T = 40$}, $N_t = 10^{28}/\rm{m^3}$, $\kappa =10^{-9}\ \rm{V\cdot m^3/C}$. 

\begin{figure}
    \includegraphics[width=0.9\linewidth]{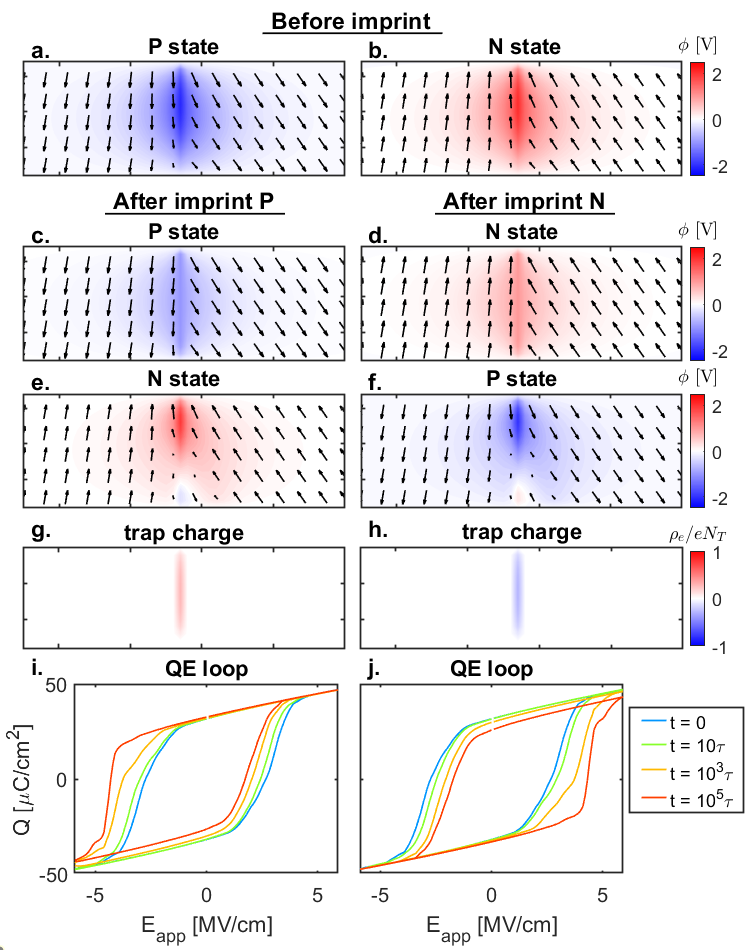}
    \caption{(a)-(f) The polarization field and electric potential before and after imprint for different polarization states; (g)(h) the trap charge density after imprint; (i)(j) the shift of hysteresis loop during the imprint process. }
    \label{fig2}
\end{figure}

Fig.2 illustrates the physical mechanisms of the \rv{GBCS} bulk imprint model. Before imprint (no trap charge yet, Fig.2(a)(b)), the strong electric field near grain boundaries, arising from the nonzero bound charge, can intrigue domain nucleation and governs the coercive field. During the imprint process, electrons/holes are trapped primarily on grain boundaries (Fig.2(g)(h)), driven by the attraction of bound charges. After imprint, for the same poling state (Fig.2(c)(d)), the local electric field is weakened by the trapped charges, increasing the energy barrier to switch to the other state. For the opposite state (Fig.2(e)(f)), a self-regulating feedback mechanism occurs: the rise in the local electric field enlarges the reversed domains, which subsequently attenuates the field itself. This domain nucleation ultimately lowers the energy barrier required for switching.   

\rv{Fig. 2(i)(j) depict the temporal evolution of hysteresis loops. After imprinting of the N (P) state, the loops shift right (left), exhibiting the asymmetry $|\Delta E_c^+|>|\Delta E_c^-|$ ($|\Delta E_c^+|<|\Delta E_c^-|$), consistent with experimental trends \cite{kim_imprint_2024}. This asymmetry arises because the assistance provided by GBCS to reverse back the polarization state is limited by the domain wall motion speed after large nuclei have formed.}

\begin{figure}
    \includegraphics[width=0.9\linewidth]{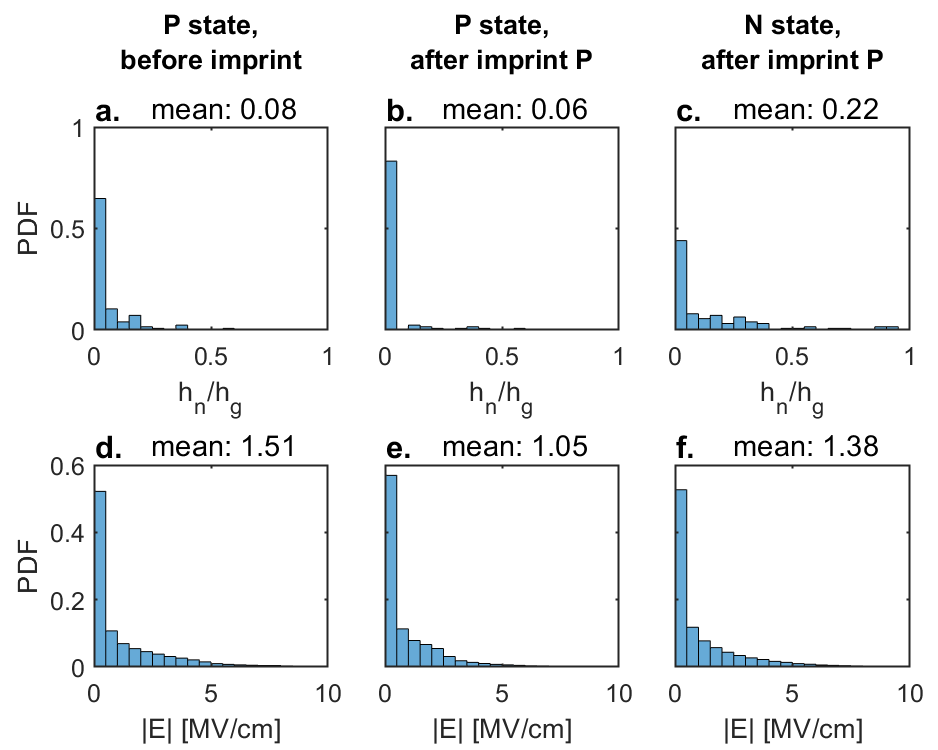}
    \caption{The probability distribution of the size of the domain nucleus (a,b,c) and the strength of local electric field (d,e,f) for the base case of the bulk imprint model. }
    \label{fig3}
\end{figure}

\begin{figure}
    \includegraphics[width=0.9\linewidth]{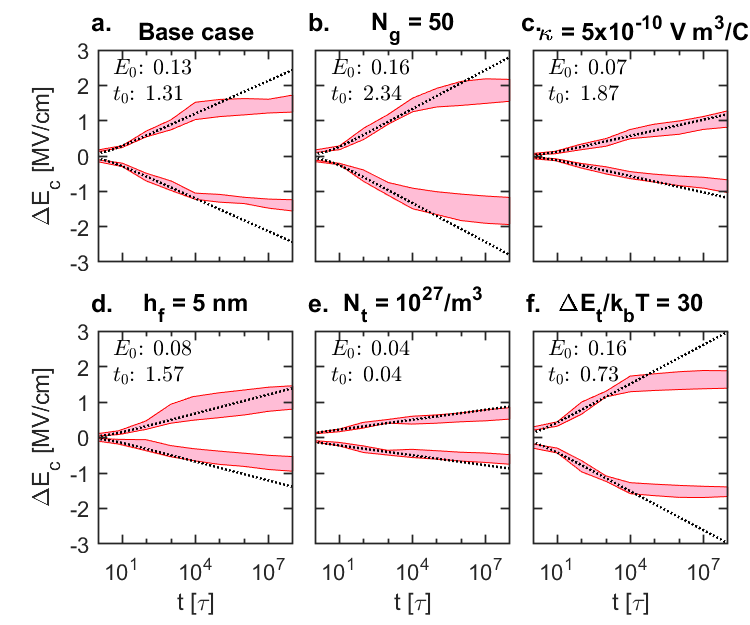}
    \caption{\rv{Temporal evolution of coercive field shift ($\Delta E_c$) during imprint of P state ($\Delta E_c<0$) and N state ($\Delta E_c>0$),  for six cases. The shaded area representing the standard deviation for 5 random polycrystal systems, while the dotted lines show the fitting  using $\Delta E_c = E_0 \ln(1+t/t_0)$. The units for $E_0$ and $t_0$ are MV/cm and $\tau$, respectively. } }
    \label{fig:fig4}
\end{figure}

To demonstrate the generality of the microscopic mechanisms illustrated in Fig.2 across polycrystalline systems, Fig.3 presents the statistical distributions of domain nucleus size $h_n$ and local electric field strength $|E|$. A domain nucleus is defined as the reversed domain (exhibiting a $P_y$ orientation opposite to that of the overall film) located between the centers of two adjacent grains. The results in Fig.3 reveal that after imprint, domain nuclei tend to shrink in the same state but expand in the opposite state. Meanwhile, the electric field strength decreases significantly in the same state, and does not change much in the opposite state (due to the self-regulating feedback mechanism discussed before). 


\rv{Finally, Fig.4 presents the time evolution of $\Delta E_c$ (averaged from $\Delta E_c^+$ and $\Delta E_c^-$)  for six parameter sets, along with the fitting curves based on $\Delta E_c = E_0 \ln(1+t/t_0)$. The early-time data are well fitted by this logarithmic law, and the value of $E_0 h_f$ is consistent with experimental measurements (0.04$\sim$0.14 V in \cite{vishnumurthy_ferroelectric_2024}). At longer times, $\Delta E_c$ saturates, a phenomenon potentially absent in experiments due to limited measurement duration. Furthermore:}


 
(1) \rv{$E_0$} increases with finer grains (Fig.\ref{fig:fig4}(a)(b)), as domain nucleation becomes more important than domain wall motion for switching. This correlation is absent in the \rv{ICS} models. 

(2) \rv{$E_0$} decreases with smaller $\kappa$ (Fig.\ref{fig:fig4}(a)(c)), because nucleation is easier as  domain wall energy decreases (smaller $\kappa$) \cite{merz_switching_1956}. This correlation is also missing in the \rv{ICS} models. Note that the same $\kappa$ may yield  larger \rv{$|\Delta E_c|$}  
in a 3D model than in a 2D model. 

(3) \rv{$E_0$} decreases in thinner films (Fig.\ref{fig:fig4}(a)(d)), possibly due to easier nucleation in confined geometry. However, thinner films usually have smaller grains \cite{thompson_grain_1990}, which can lead to a larger imprint effect. Therefore, to describe the scaling effect using the bulk imprint model, grain statistics information is required. In comparison, the \rv{ICS} models predict that thinner films have a stronger imprint effect, assuming that the dead layer does not change. Concrete experimental data for the scaling effect of \rv{$\Delta E_c$} are still deficient. 

\rv{(4) $E_0$ decreases with increasing $N_t$, as expected.}

\rv{(5) $E_0$ increases while $\tau_0$ decreases as $\Delta E_t/k_b T$ decreases, confirming the same trend as the experimental temperature effect in \cite{vishnumurthy_ferroelectric_2024}. A comprehensive model for temperature dependence will need to incorporate the changes in $f_{\rm{Landau}}$ and $f_{\rm{grad}}$, a task we leave for future studies.} 


\section{Conclusion and perspectives}


\rv{To conclude, we propose a bulk imprint mechanism—grain-boundary charge screening (GBCS)—for polycrystalline ferroelectric films. While acknowledging the role of interface charge screening (ICS), the contribution of GBCS is critical due to the intense local electric fields (on the order of MV/cm) near grain boundaries, which facilitate domain nucleation and govern polarization switching. The resulting imprint behavior is explained by the GBCS-mediated tuning of the domain nucleation energy barrier. Our GBCS model aligns with key experimental features, including the logarithmic time evolution, thermal acceleration, and asymmetric shift of hysteresis branches. Further validation requires comprehensive experimental data on the effects of material properties (e.g., domain wall energy), polycrystal statistics, film thickness, and interface engineering. Moreover, the GBCS framework may be extended to explain the fatigue process: the domains that do not participate in the switching cycles can be pinned on grain boundaries, inducing subsequent pinning of neighboring domains. This work provides new insights into the mechanisms of imprint, and establishes a new mathematical framework for investigating related reliability issues in ferroelectrics. }

\begin{acknowledgments}
This work was supported in part by the NSFC under Grants  92164204, 62222119.
\end{acknowledgments}


\bibliography{my_bib}

\appendix

\end{document}